\documentclass{IEEEtran}
\usepackage{array}
\usepackage{graphicx, subfigure}
\usepackage{float}
\usepackage{multirow}
\usepackage{rotating}
\usepackage{algorithmic,algorithm}
\usepackage{amsmath} 
\usepackage{amsfonts}
\usepackage{amssymb}  
\usepackage{amsthm}
\usepackage{bm}

\begin{document}
\title{Achieving proportional fairness with a control theoretic approach in error-prone 802.11e WLANs}
\author{Xiaomin Chen and Shuang-Hua Yang\IEEEauthorrefmark{1}
\\Department of Computer Science, Loughborough University
\thanks{\IEEEauthorrefmark{1} corresponding author S.H.Yang@lboro.ac.uk}}
\maketitle

\begin{abstract}
This letter proposes a control theoretic approach to achieve proportional fairness amongst access categories (ACs) in an error-prone EDCA WLAN for provision of distinct QoS requirements and priority parameters. The approach adaptively adjusts the minimum contention window of each AC to derive the station attempt probability to its optimum which leads to a proportional fair allocation of station throughputs. Evaluation results demonstrate that the proposed control approach has high accuracy performance and fast convergence speed for general network scenarios.
\end{abstract}

\section{Introduction}

802.11e Enhanced Distributed Channel Access (EDCA) extends the basic Distributed Coordination Function (DCF) by classifying traffic into different Access Categories (ACs) to provide service differentiation.  Traffic with higher QoS requirements, e.g. shorter delay deadline,  is assigned a higher priority, and thus, on average, waits for less time before being sent to the channel. Although it provides QoS enhancement, EDCA is essentially unfair because it benefits high-priority traffic at the cost of low-priority traffic's performance. If a network is loaded with a large proportion of high-priority flows, the channel will nearly be all occupied by high-priority flows, e.g. VoIP or video streaming flows, low-priority traffic however , such as email or web browsing data, will suffer severe starvation. Therefore, it is of significance to seek for a fair allocation of network resources (e.g. throughput, airtime and etc.) amongst different traffic types, and meanwhile, guarantee the specific QoS requirements and service differentiation.

Recently~\cite{Xiaomin2} derives the proportional fair allocation amongst different ACs in an 802.11e EDCA WLAN given distinct average delay deadlines, and proposes a control theoretic method to achieve it. In this letter we extend that work in a number of directions. Firstly, the analysis in~\cite{Xiaomin2} assumes that transmissions are error-free while in this letter we relax this assumption. Packet losses are caused by either collisions or channel errors. Secondly, in~\cite{Xiaomin2} we fix $CW_{max}=CW_{min}$ while in this letter due to the assumption of error-prone channels, packets need to be retransmitted and hence we set $CW_{max}=2^m\cdot CW_{min}$ and $m$ is the retry limit. Thirdly, the retransmission delay in this letter is not simply the collision delay as considered in~\cite{Xiaomin2}, it should also include the delay caused by retransmission(s) of a TXOP burst which is determined by the packet error rate.

\section{Proportional fairness in 802.11e EDCA WLANs}\label{Sec:Fairness}

\subsection{Network model}
We consider a single-hop 802.11e EDCA WLAN with one AP and $n$ client stations. Traffic flows are classified into $N$ ACs. We assume that each station only transmits traffic of a single AC, so no virtual collisions  are considered in our analysis. The number of stations in the $i$th AC is $n_i$.
 Stations in the $i$th AC are assumed to have the same packet error rate $p_i$, and transmit packets at the same PHY data rate $r_i$. The packet size of all ACs is $L$ bits.
\subsection{Station throughput}
We start with the analysis of station throughput under saturated network loads. The following parameters are defined for AC $i$:
\begin{itemize}
  \item $CW_{min}^i$ is the minimum contention window;
  \item $CW_{max}^i$ is the maximum contention window;
  \item $AIFS_i$ is the duration of the Arbitration Inter-Frame Space, i.e. $AIFS_i=SIFS+t_i\times\sigma$, in which $t_i$ is the number of included time slots, $SIFS$ is the duration of the Short Inter-Frame Space and $\sigma$ is the duration of a physical time slot;
  \item $M_i$ is the number of packets contained in a TXOP burst.
\end{itemize}

Due to the use of TXOP bursting, the RTS/CTS exchange mechanism is used to make fast recovery from collisions.
If any packet within the TXOP burst fails due to channel errors, the burst is terminated and the station contends again for the channel to retransmit the failed packet. Therefore, a MAC time slot may be a PHY idle slot, a TXOP transmission or a RTS/CTS transmission.

Let $\tau_i$ denote the probability that a station in AC $i$ attempts to transmit in a time slot, and $\boldsymbol{\tau}=\{\tau_i, \ i=0,\cdots,N-1\}$. The station throughput of AC $i$ is given by
\begin{equation}\label{thrpt}\begin{split}
&s_{i}(\boldsymbol{\tau})=\\&\frac{P_i^{T}\mathbb{E}(Pld_i^{txop})}{{P^{I}}\sigma+\sum\limits_{i=0}^{N-1}n_iP_i^{T}\mathbb{E}(T_i^{txop})+(1-P^{I}-\sum\limits_{i=0}^{N-1}n_iP_i^{T})T^{col}}
\end{split}\end{equation}
in which $P^{I}=\prod\limits_{i=0}^{N-1}(1-\tau_i)^{n_i}$ is the probability of an idle time slot
and $P_i^{T}=\tau_{i}(1-\tau_i)^{n_i-1}\prod\limits_{j=0,j\neq i}^{N-1}(1-\tau_j)^{n_j}$ is the probability that a station of AC $i$ makes a TXOP transmission.
$T^{col} = T_{RTS}+EIFS$ is the duration of a collision, where $EIFS$ is the duration of the Extended Inter-Frame Space, {which is given by $EIFS=T_{ACK} + SIFS + DIFS$}. $\mathbb{E}(T_i^{txop})$ is the expected duration of a TXOP transmission from a station of AC $i$,
\begin{equation*}\begin{split}
  \mathbb{E}(T_i^{txop})=&\sum\limits_{k=1}^{M_i}(1-p_i)^{k-1}p_i\left(k\Big(\frac{L}{r_i}+T^{oo}\Big)+T_i^o\right)\\&+(1-p_i)^{M_i}\left(M_i\Big(\frac{L}{r_i}+T^{oo}\Big)+T_i^o\right)
\end{split}\end{equation*}
where $T^{oo}=T_{PHYhdr}+2SIFS+T_{ACK}$ is the protocol overheads associated with a single packet transmission within a TXOP burst, and $T_i^o=T_{RTS}+SIFS+T_{CTS}+AIFS_i$ is the protocol overheads associated with the transmission of the burst. $\mathbb{E}(Pld_i^{txop})$ is the expected payload transmitted in a TXOP burst of AC $i$,
\begin{equation*}
  \mathbb{E}(Pld_i^{txop})=\sum\limits_{k=1}^{M_i-1}(1-p_i)^{k}p_ikL+(1-p_i)^{M_i}M_iL
\end{equation*}

\subsection{Average delay}
Next, we will calculate the average packet transmission delay. As the delay is mainly dominated by the number of retransmission attempts caused by either collisions or erroneous transmissions, we ignore the countdown and blocking delays considered in~\cite{Xiaomin2}. The average delay of a TXOP burst consists of the following two expected delays:
\begin{itemize}
  \item  \emph {Expected collision delay}:
  Assuming that $m$ is the retry limit and $CW_{max}^i= 2^m{CW}_{min}^i$, the expected collision delay is
       \begin{equation*}
         D_i^{col}=T^{col}
         \bigg(\sum\limits_{j=1}^{m}j\big(P_i^{C}\big)^{j}\big(1-P_i^{C}\big)+(m+1)\big(P_i^{C}\big)^{m+1}\bigg)
       \end{equation*}
        in which $P_i^{C}$ is the conditional collision probability,
   \begin{equation*}
     P_i^{C}=1-(1-\tau_i)^{n_i-1}\prod\limits_{\substack{j=0,j\neq i}}^{N-1}(1-\tau_j)^{n_j}
   \end{equation*}
  \item \emph {Expected TXOP transmission delay}:
  The expected TXOP transmission delay is calculated by multiplying the expected number of transmissions by the expected duration of a TXOP burst.
  \begin{equation*}\begin{split}
    D_i^{txop}=&\mathbb{E}({T}_i^{txop})\bigg(\sum\limits_{j=1}^{m+1}j\big(1-P_i^{C}\big)^{j}\big(P_i^{E}\big)^{j-1}\big(1-P_i^{E}\big)\\&+(m+1)\big(1-P_i^{C}\big)^{m+1}\big(P_i^{E}\big)^{m+1}\bigg)
  \end{split}\end{equation*}
  in which $P_i^{E}=1-(1-p_i)^{M_i}$ is the probability of a TXOP failure due to channel noise.

\end{itemize}
Combining the above two delays, the average delay of a single packet within a TXOP burst is therefore given by
\begin{equation}\label{delay}
  D_i=\frac{D_i^{col}+D_i^{txop}}{\sum\limits_{k=1}^{M_i}k(1-p_i)^{k-1}p_i+M_i(1-p_i)^{M_i}}
\end{equation}

\subsection{Relationship between $\tau_i$ and ${CW_{min}^i}$}
The station throughput and average delay derived above are functions of station attempt probabilities. Next we will derive the relationship between $\tau_i$ and ${CW_{min}^i}$ based on the throughput model proposed in~\cite{Kosek} but consider error-prone channels. The probability that a transmission fails due to either collisions or channel noise is given by
\begin{equation}\label{popt}
  P_i^{F}=1-\big(1-P_i^{C}\big)\big(1-P_i^{E}\big)
\end{equation}
The probability that the backoff counter is suspended due to a busy channel during the period of $AIFS_i$ is
\begin{equation*}
  P_i^{B}=1-\Big[(1-\tau_i)^{n_i-1}\prod\limits_{\substack{j=0\\j\neq i}}^{N-1}(1-\tau_j)^{n_j}\Big]^{t_i-t_{min}+1}
\end{equation*}
in which $t_{min}$ is the minimum $t$ value among all ACs.
The relationship between $\tau_i$ and $CW_{min}^i$ is then given by
\begin{equation}\label{CW}\begin{split}
  &\tau_i=\\&\left(1-\frac{1}{2\big(1-P_i^{B}\big)}+\frac{CW_{min}^i\left(1-P_i^{F}\right)\big(1-(2P_i^{F})^{m+1}\big)}{2\left(1-P_i^{B}\right)\left(1-2P_i^{F}\right)\big(1-(P_i^{F})^{m+1}\big)}\right)^{-1}
\end{split}\end{equation}

\subsection{Proportional fair allocation}\label{Sec:Fairness}

In this section we aim at finding the optimal $\boldsymbol{\tau}$ to achieve proportional fair allocation of station throughputs given distinct average delay constraints for different ACs.

\subsubsection{Utility function}
The utility function is defined as the sum of the log of station throughputs，
\begin{equation*}\begin{split}
&\underset{\boldsymbol{\tau}}
{\text{max}}\quad U(\boldsymbol{\tau}):= \sum\limits_{i=0}^{N-1}n_i\log s_i(\boldsymbol{\tau})\\
&\text{s. t. } \quad D_i(\boldsymbol{\tau})\leq d_i \quad 0\leq i \leq N-1,\\
&\quad \quad \quad 0 < \tau_i < 1 \quad   \ 0\leq i \leq N-1.
\end{split}
\end{equation*}
in which $d_i$ is the delay deadline for AC $i$. The station throughput is given by Eqn.~(\ref{thrpt}), and the average delay is given by Eqn.~(\ref{delay}).
\subsubsection{Solving the optimisation problem}
We follow the algorithm described in~\cite{Xiaomin2} to solve this optimisation problem. The steps are summarised as below:

(i) Convert to a convex optimisation problem by making a log transformation $\eta_i=\log\frac{\tau_i}{1-\tau_i}$;

(ii) Introduce Lagrange multipliers using Karush-Kuhn-Tucker (KKT) conditions;

(iii) Calculate the optimal Lagrange multipliers using a standard sub-gradient approach;

(iv) Calculate the optimal $\boldsymbol{\tau}$ by plugging the optimal Lagrange multipliers back to the Lagrangian.

\section{Centralised closed-loop control approach}

In this section, we design a centralized adaptive control approach to implement the desirable proportional fairness in real networks.  Based upon the analysis in Section~\ref{Sec:Fairness}, the proportional fairness is achieved when the station attempt probability reaches its optimum value. The variable $\boldsymbol{\tau}$ is only determined by $CW_{min}$ with $CW_{max}$, $AIFS$ and $TXOP$ taking the recommended values. Our approach uses a multi-variable closed-loop control system to tune $\boldsymbol{CW_{min}}=\{CW_{min}^i, \ i=0,\cdots,N-1 \}$  so as to drive the station attempt probability to its optimum. As the station attempt probability is hard to measure in real networks, we measure the transmission failure probability $\boldsymbol{p^{F}}(\boldsymbol{\tau})=\{p_i^{F}(\boldsymbol{\tau}), \ i=0,\cdots,N-1 \} $ instead of $\boldsymbol{\tau}$. The observed failure probability can be estimated as $p_i^{F}=\frac{N_i^1}{N_i^0+N_i^1}$, in which $N_i^1$ represents the number of packets of AC $i$ that are received by the AP within a beacon interval with the retry bit equal to $1$, and $N_i^0$ represents the number of received packets with the retry bit set to $0$. The estimation is based on the assumption that all packets will be eventually received within the retry limit. 
\subsection{Adaptive control algorithm}

 The adaptive control algorithm is performed centrally at the AP every beacon interval, which is typically 100ms in 802.11 WLANs, and consists of two steps:
 \begin{enumerate}
   \item At the end of each beacon interval, the observed failure probability $\boldsymbol{p^{F}}$ resulting from the current $\boldsymbol{CW_{min}}$ is measured by the AP.
   \item The controller computes a new set of $\boldsymbol{CW_{min}}$ based on the measured $\boldsymbol{p^{F}}$.  An update of $\boldsymbol{CW_{min}}$ is then distributed to the WLAN by the next beacon frame.
 \end{enumerate}



\subsection{Linearisation of the non-linear plant}

As the period of the adaptive control algorithm is long enough to assume that the measurement corresponds to stationary conditions, i.e. the system has no memory, $\boldsymbol{p^{F}}$ depends only on the current $\boldsymbol{CW_{min}}$.
Eqn.~(\ref{popt}) and Eqn.~(\ref{CW}) give a non-linear relationship between $\boldsymbol{p^{F}}$ and $\boldsymbol{CW_{min}}$.
To simplify the controller design, we will use a linear approximation to this non-linear relationship around the stable point of operation,
\begin{equation}\label{output}
  \boldsymbol{p^{F}}-\boldsymbol{p^{F*}}=(\boldsymbol{CW_{min}}-\boldsymbol{CW_{min}^*})\cdot \boldsymbol{H}
\end{equation}
in which
\begin{equation*}
  \boldsymbol{H}=\left(
     \begin{array}{cccc}
       \frac{\partial p^{F}_0}{\partial CW_{min}^0} & \frac{\partial p^{F}_1}{\partial CW_{min}^0} & \cdots & \frac{\partial p^{F}_{N-1}}{\partial CW_{min}^0} \\
       \frac{\partial p^{F}_0}{\partial CW_{min}^1} & \frac{\partial p^{F}_1}{\partial CW_{min}^1} & \cdots & \frac{\partial p^{F}_{N-1}}{\partial CW_{min}^1}  \\
       \vdots & \vdots & \ddots & \vdots \\
        \frac{\partial p^{F}_{0}}{\partial CW_{min}^{N-1}} & \frac{\partial p^{F}_{1}}{\partial CW_{min}^{N-1} }& \cdots & \frac{\partial p^{F}_{N-1}}{\partial CW_{min}^{N-1}} \\
     \end{array}
   \right)
\end{equation*}

\subsection{State feedback control }

 With the linearisation, the WLAN can be represented as  a discrete MIMO state-space model. The failure probability at instant $k+1$ is determined by the minimum contention window input to the WLAN at instant $k$, the state and measurement equations are therefore given by
\begin{equation*}\left\{
                   \begin{array}{ll}
                     \boldsymbol{x}(k+1)=\boldsymbol{B}\boldsymbol{u}(k)\\
                     \boldsymbol{y}(k)=\boldsymbol{C}\boldsymbol{x}(k)
                   \end{array}
                 \right.
\end{equation*}
 in which the system state is $ \boldsymbol{x}(k)=[\boldsymbol{p^{F}}(k)]^T$; the system input is $\boldsymbol{u}(k)=[\boldsymbol{CW_{min}}(k)]^T$; the system model matrices are $\boldsymbol{B}=-\boldsymbol{H}^T$
  and $\boldsymbol{C}=\boldsymbol{I}_{N\times N}$;
the system output is $\boldsymbol{y}(k)=\boldsymbol{C}\boldsymbol{x}(k)=[\boldsymbol{p^{F}}(k)]^T$.

\begin{figure}
  \begin{center}
  \includegraphics[height=2cm, width=\columnwidth]{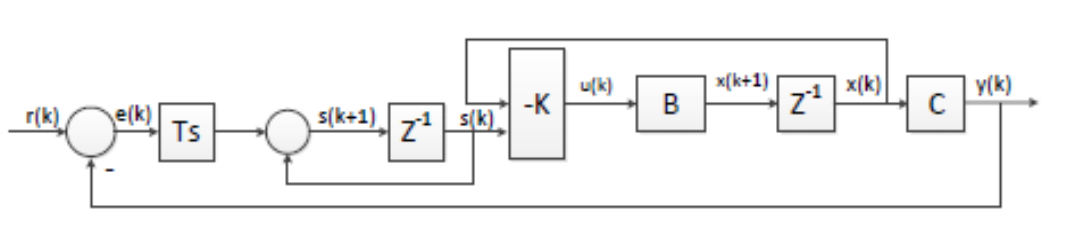}\\
   \caption{LQI Controller}\label{Fig：LQI}
  \end{center}
\end{figure}

 The control task can be accomplished by using the LQI control method ~\cite{LQI} to design our controller. Fig.~\ref{Fig：LQI} shows the control block diagram for the system, in which the controller input ${r}(k)={p^F}^*$ is the optimal failure probability, $e(k)=r(k)-y(k)$ is the error between the output feedback and the input reference value, and  ${s}(k)={s}(k-1)+T_se(k-1)$ is the output of a discrete integrator and $T_s$ is the sampling period. The  $\boldsymbol{K} \in \mathbb{R}^{N\times 2N}$ is the control gain matrix.

The LQI controller computes an optimal state-feedback control law
\begin{equation*}
  \boldsymbol{u}(k)=-\boldsymbol{K}\boldsymbol{z}(k)
\end{equation*}
by minimising the quadratic cost function
\begin{equation*}
  J(\boldsymbol{u}(k))=\sum\limits_{k=0}^{\infty}\big(\boldsymbol{z}^T(k)\boldsymbol{Q}\boldsymbol{z}(k)+\boldsymbol{u}^T(k)\boldsymbol{R}\boldsymbol{u}(k)\big)
\end{equation*}
for any initial state $\boldsymbol{x}(0)$, in which $\boldsymbol{z}(k)=[\boldsymbol{x}(k);\boldsymbol{s}(k)]$. The optimal state feedback gain matrix $\boldsymbol{K}$  is computed by solving the associated discrete algebraic Riccati equation~\cite{LQI}.

The matrices $\boldsymbol{Q}$ and $\boldsymbol{R}$ are the weighting matrices respectively indicating the state and control cost penalties. $\boldsymbol{Q}$ and $\boldsymbol{R}$ are required to be real symmetric and positive definite.  The selection of weighting matrices $\boldsymbol{Q}$ and $\boldsymbol{R}$ affect the convergence speed of the controller.


\section{Performance evaluation}
\subsection{Impact of packet error rates}
\begin{figure}
  \begin{center}
  \includegraphics[height=4cm,width=0.8\columnwidth]{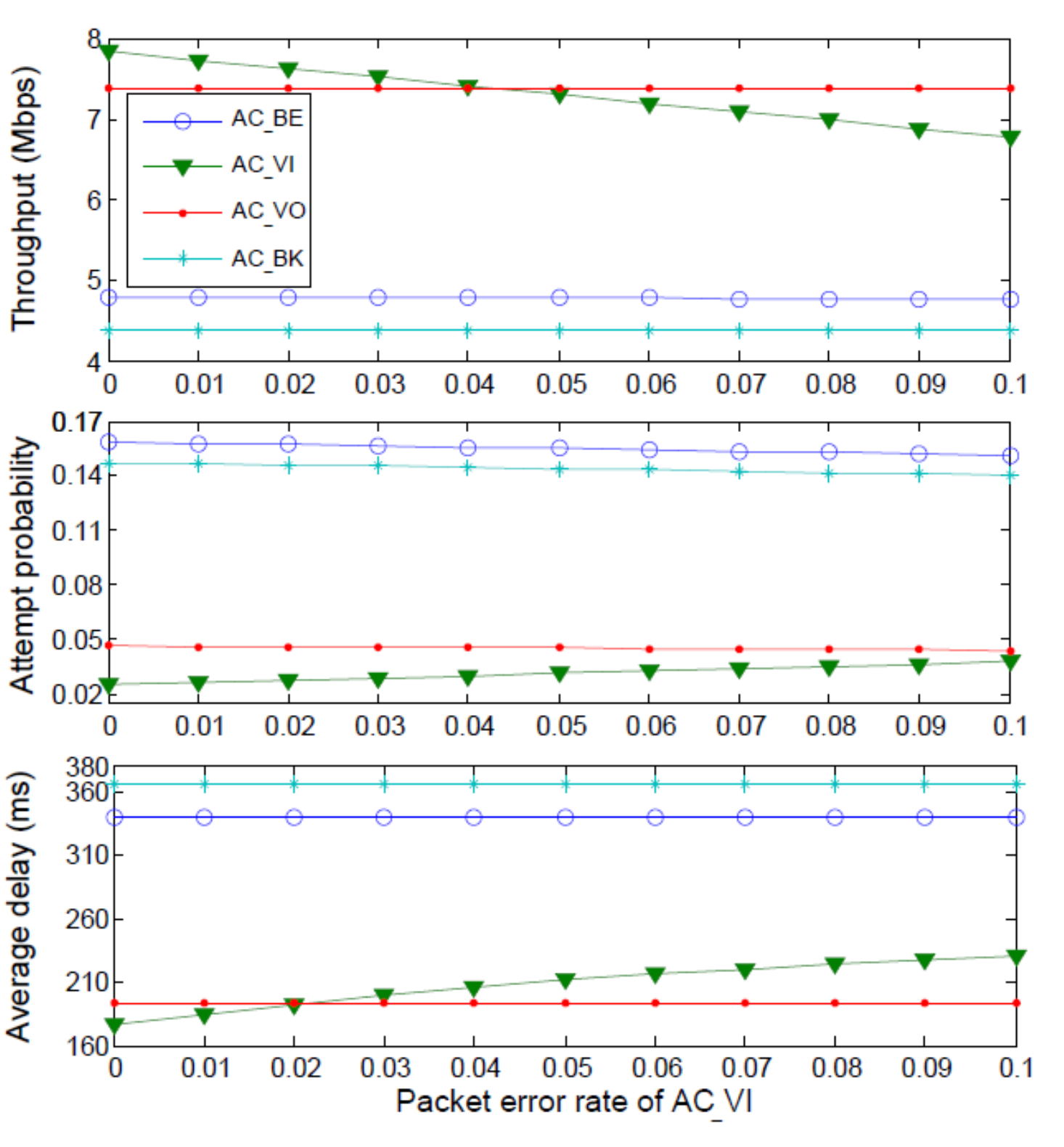}\\
   \caption{Delay and throughput performances, $L=8000$ bits,  $n_{BE}=n_{VI}=n_{VO}=n_{BK}=1$, $r_{BE}=r_{VI}=r_{VO}=r_{BK}=54$Mbps, $p_{BE}=p_{VO}=p_{BK}=10^{-6}$, $d_{BE}=400$ms, $d_{VI}=250$ms, $d_{VO}=200$ms and $d_{BK}=400$ms. The delay deadlines are not reached in this example.}\label{Fig:PER}
  \end{center}
\end{figure}

We will first evaluate the impact of packet error rates on the station throughput and delay performances. Consider a WLAN with four stations and each station transmits a traffic flow of a distinct AC.
 The protocol parameter values used in the simulations are listed in Table~\ref{protocol_para}. Fig.~\ref{Fig:PER} plots the throughput, average delay and station attempt probability of four ACs as the packet error rate $p_{VI}$ varies while keeping $p_{BE}=p_{VO}=p_{BK}=10^{-6}$.
 It can be seen that as the PER increases, the station in AC$\_$VI attempts harder with an increasing attempt probability but the throughput and delay performances irreversibly degrade. The increase of $p_{VI}$ also results in the decrease of attempt probabilities of the other three ACs, but the coupling of station transmissions lead to invariant delay and throughput performances, which implies that the delay and throughput of an AC are only dependent on its own PER.
 \begin{table}\tiny
\caption{802.11 protocol parameters used in the simulations}\label{protocol_para}
\centering
\begin{tabular}{|c c| c c |c c |}
  \hline
   $\sigma$    & 9  $\mu$s  & $T_{PHYhdr}$     & 20  $\mu$s     & $SIFS$      & 16 $\mu$s    \\    \hline
  $DIFS$       & 34 $\mu$s  & $T_{CTS}$        & 38.67 $\mu$s   & $EIFS$      & 88.67 $\mu$s \\    \hline
    $T_{RTS}$   & 46.67$\mu$s &$T_{ACK}$      & 38.67 $\mu$s   & $AIFSN$      & $[7,3,2,2]$     \\  \hline
 \end{tabular}
\end{table}
\subsection{Air-time}
\begin{figure}
  \centering
  \subfigure[Flow total air-time]{\label{Fig:airtime}\includegraphics[height=3.15cm, width=0.49\columnwidth]{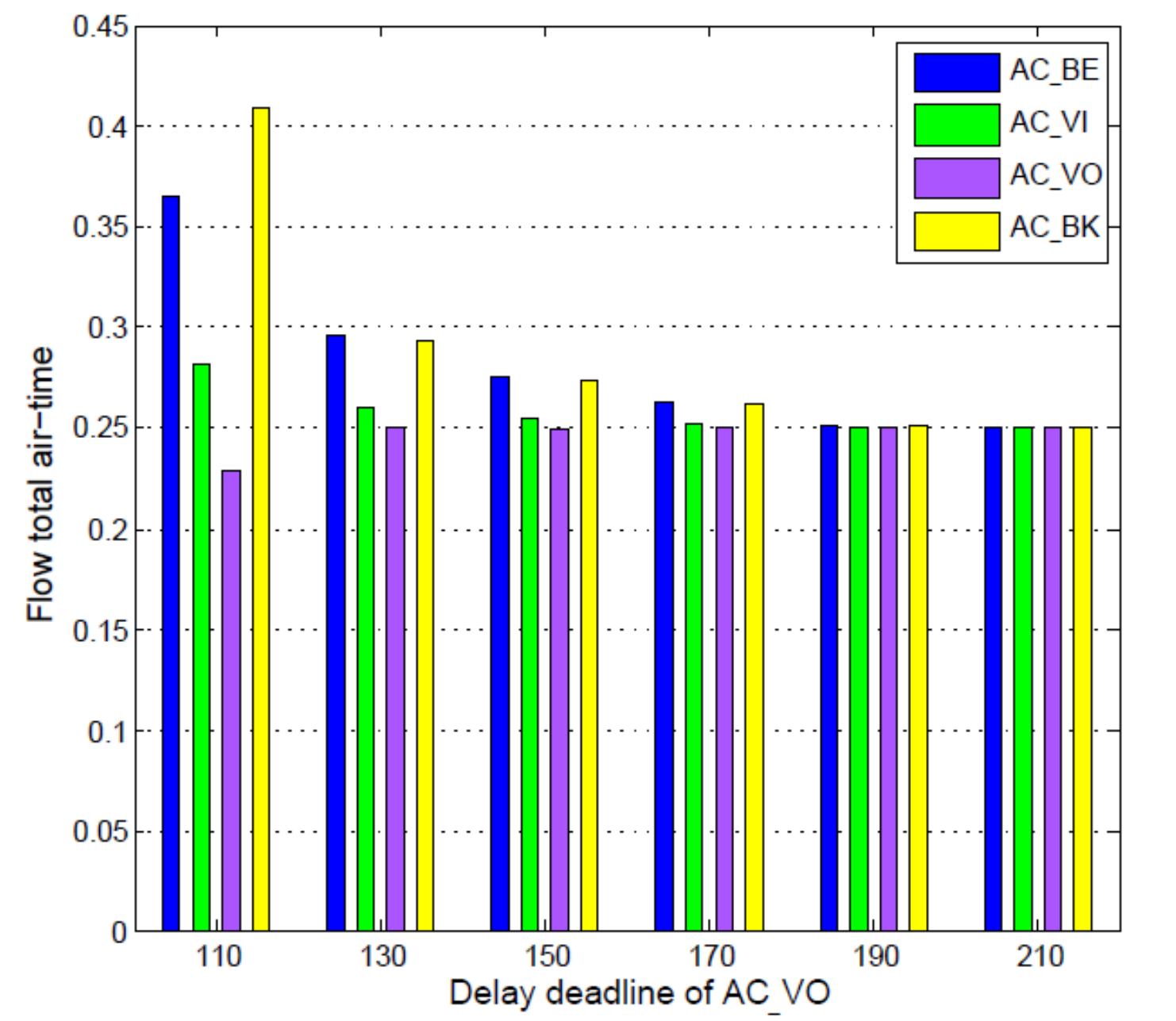}}
  \subfigure[Sum of air-times]{\label{Fig:SumAirtime}\includegraphics[height=3.15cm, width=0.49\columnwidth]{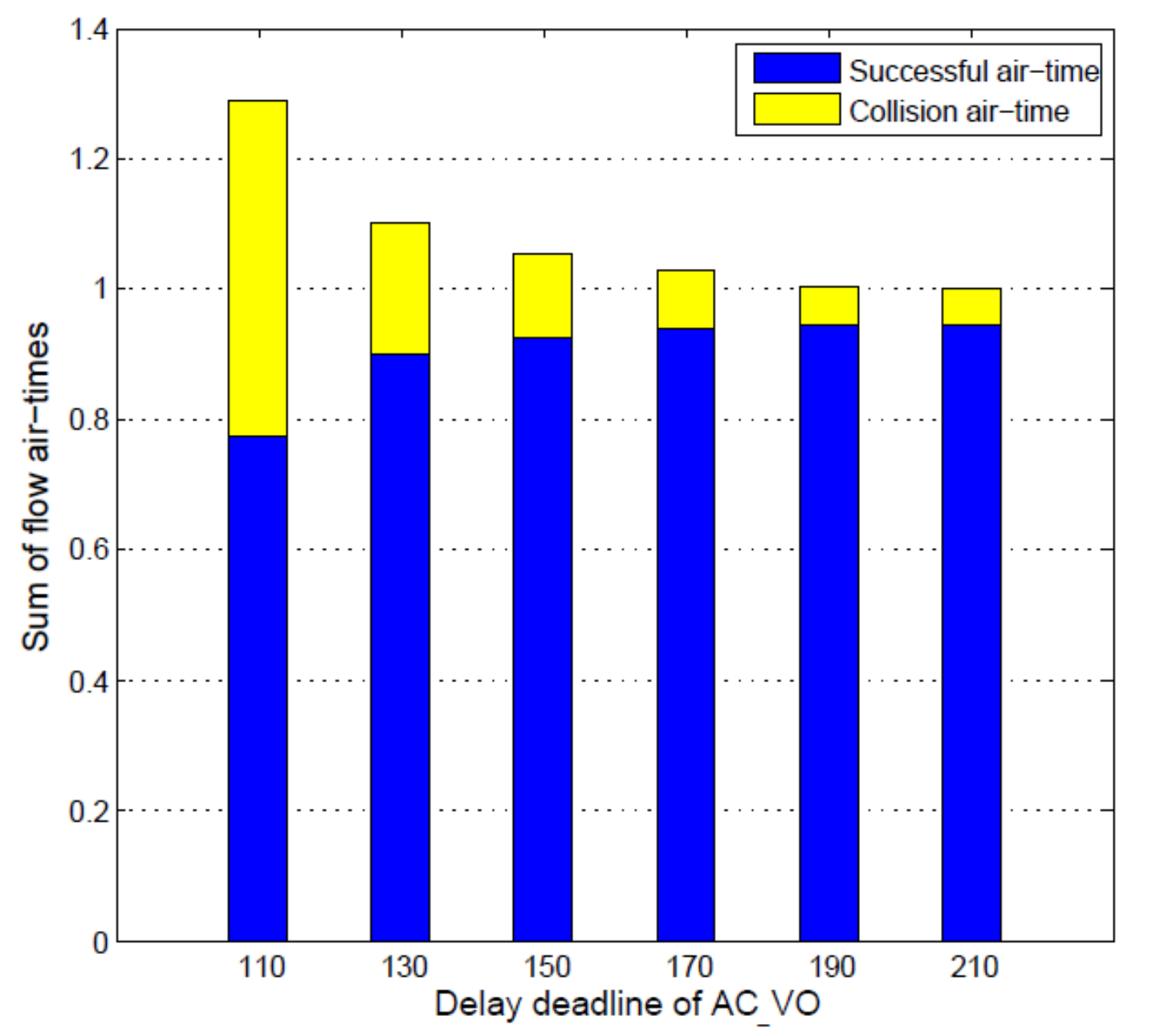}}
  \caption{ Air-time allocation, $L=8000$ bits, $n_{BE}=n_{VI}=n_{VO}=n_{BK}=1$,  $r_{BE}=r_{VI}=r_{VO}=r_{BK}=54$Mbps, $p_{BE}=p_{VI}=p_{VO}=p_{BK}=0.001$, $d_{BE}=400$ms, $d_{VI}=210$ms, $d_{BK}=400$ms.}
\end{figure}

 The \emph{flow total air-time} is defined as the fraction of time used for transmitting a flow. For a flow of AC $i$, the flow total air-time consists of the successful flow air-time and the collision air-time, given by
 \begin{equation*}
    T_i^{air}=\frac{P_i^{T}\mathbb{E}(T_i^{txop})+\tau_iP_i^{col}T^{col}}{{P^{idle}}\sigma+\sum\limits_{i=0}^{N-1}n_iP_i^{T}\mathbb{E}(T_i^{txop})+(1-P^{idle}-P^{succ})T^{col}}
 \end{equation*}
Again we consider a WLAN with four stations individually carrying a flow of a distinct AC.
Fig.~\ref{Fig:airtime} plots the flow total air-times of four ACs as the delay deadline of AC$\_$VO varies while keeping that of the other three ACs fixed.
It can be seen that when $d_{VO}\leq 210$ms, the delay constraint of AC$\_$VO keeps tight, and four flows have different air-times. As $d_{VO}$ increases, the air-time allocation is getting more and more even. When $d_{VO}$ increases up to 210ms, the four flow total air-times are equalised and all the delay constraints become loose now. Fig.~\ref{Fig:SumAirtime} plots the sum of flow total air-times.
We find that if any of the imposed delay constraints is tight, the air-times sum is larger than 1, otherwise they sum to unity. As the delay deadline is increased, the sum of collision air-times is decreasing while the sum of successful air-times is increasing. The fact that the air-times sum is not less than 1 is reasonable as the flow air-time usage overlaps due to collisions. This does not imply that the channel idle probability $P^{idle}=0$.

\subsection{Adaptivity}
We will next evaluate the adaptivity of the proposed control method to changes in the network size.  The example is depicted in Fig.~\ref{Fig:Example}.
Four stations, one in each AC, join the network at $t=0$s. The station in AC$\_$VO leaves at
t = 60s. One more station in AC$\_$BE joins the network at $t=110$s, and  the station in AC$\_$VI leaves at $t=210$s.
Fig.~\ref{Fig:AdaptivityResults} plots the variation of $CW_{min}$ and the station throughput over time. The selection of matrices {\textbf{Q} and {\textbf{R} uses the trial and error method. It can be seen that $CW_{min}$ converges to the desirable value very quickly when the network size changes as long as proper {\textbf{Q} and {\textbf{R}} are chosen. Moreover, the steady-state errors can be neglected, which means the control system has high accuracy performance.
\begin{figure}
  \begin{center}
  \includegraphics[height=1.5cm, width=0.9\columnwidth]{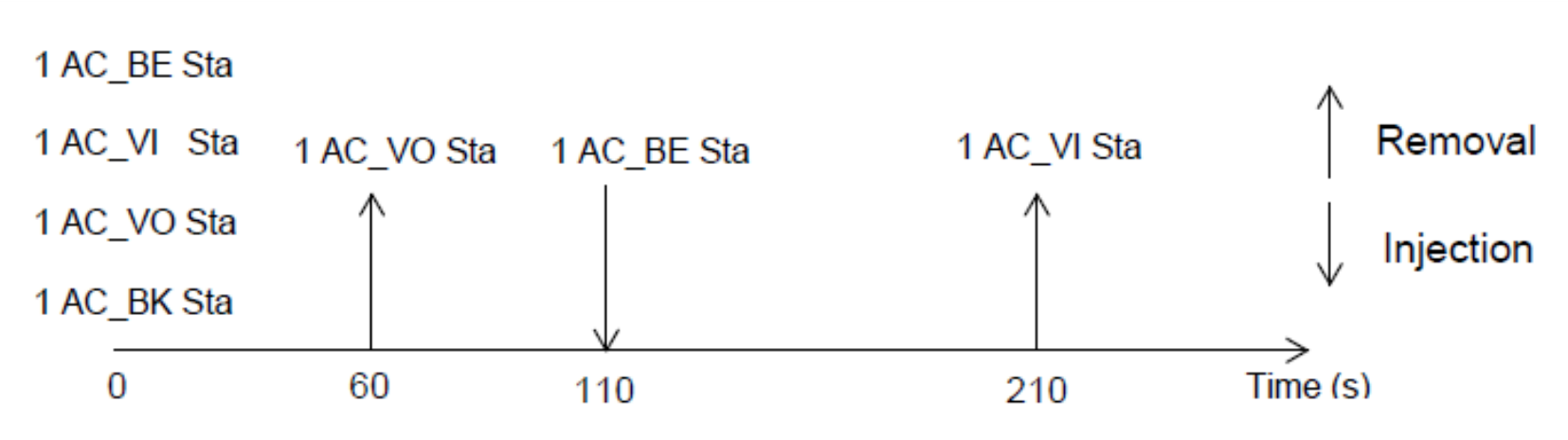}\\
   \caption{Injection and/or removal of stations in the WLAN}\label{Fig:Example}
  \end{center}
\end{figure}
\begin{figure}[!htb]
  \centering
  \subfigure[Contention window]{\label{Fig:CW}\includegraphics[height=3.15cm, width=0.49\columnwidth]{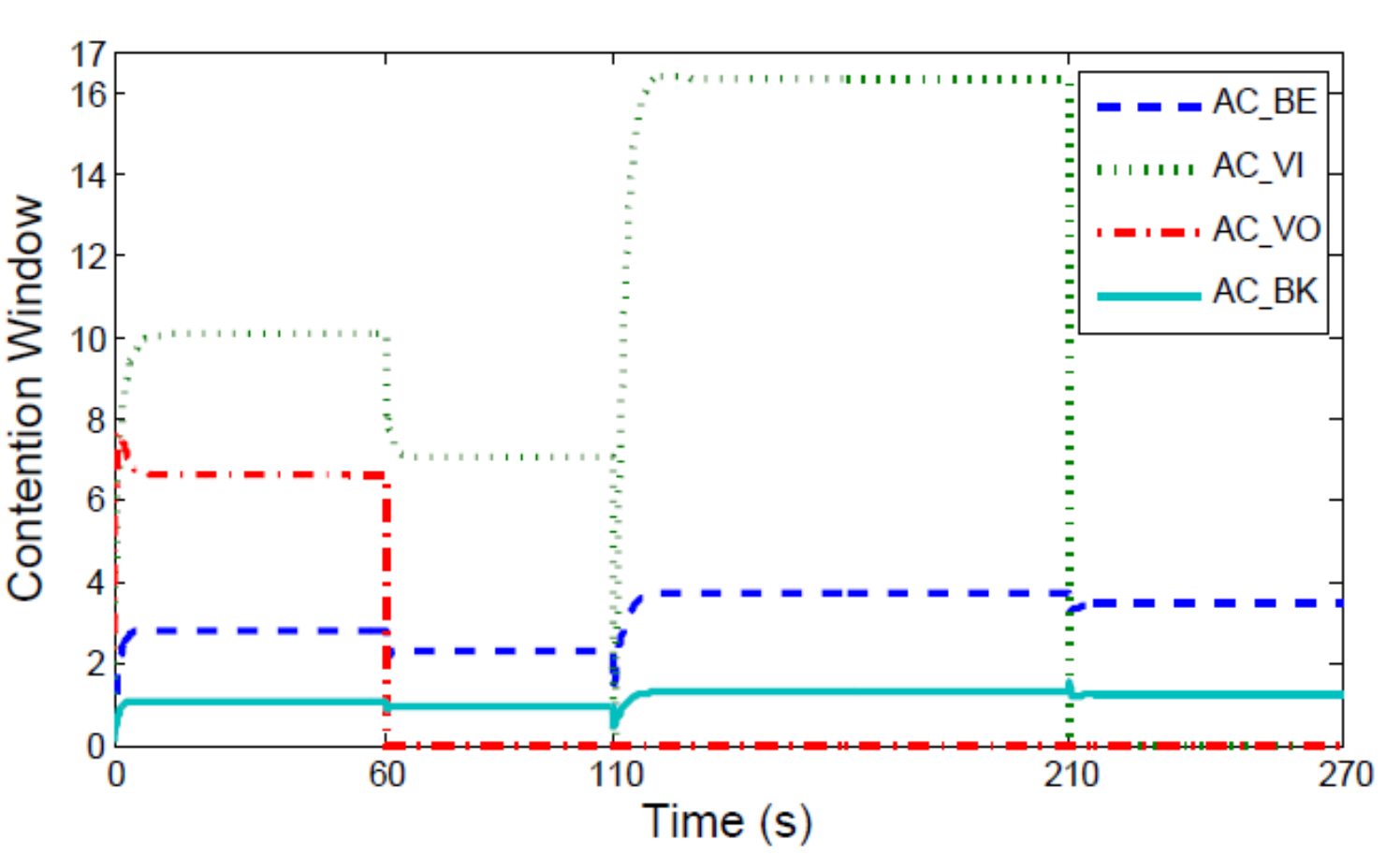}}
  \subfigure[Throughput]{\label{Fig:THRPT}\includegraphics[height=3.15cm, width=0.49\columnwidth]{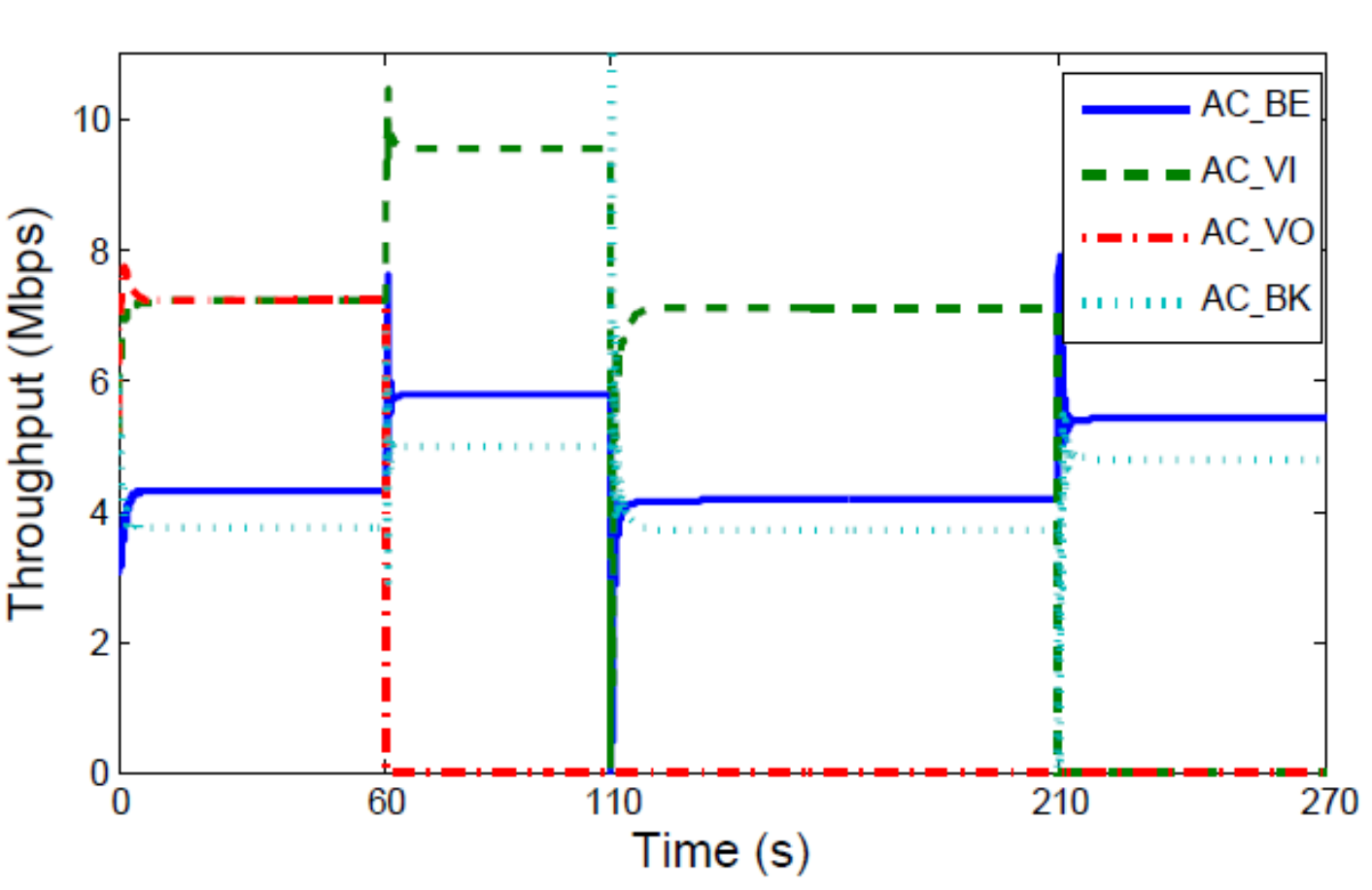}}
  \caption{ Adaptivity to changes of network size, $L=8000$ bits,  $r_{BE}=r_{BK}=36$Mbps, $r_{VI}=r_{VO}=54$Mbps, $p_{BE}=p_{BK}=0.001$, $p_{VI}=p_{VO}=0.01$, $d_{BE}=d_{VI}=d_{VO}=d_{BK}=400$ms.}\label{Fig:AdaptivityResults}
\end{figure}
\section{Conclusions}
This letter proposes a closed-loop control approach to achieve proportional fair allocation in an error-prone EDCA WLAN. The derived proportional fairness assigns unequal air-times to flows if the imposed delay constraint is tight. We have demonstrated using evaluation results that the proposed control approach has high accuracy and fast convergence speed, and is adaptive to general network scenarios.

\end{document}